\documentclass[12pt]{article}
\usepackage[margin=2cm]{geometry}
\usepackage{comment}
\usepackage{amsmath,amssymb,extarrows,mathtools,graphicx,subfigure,setspace}
\usepackage{cite}
\usepackage{epstopdf}
\usepackage{slashed}
\usepackage{color}
\usepackage{braket}
\makeatother

\newcommand{\be}{\begin{equation}}
\newcommand{\bea}{\begin{eqnarray}}
\newcommand{\eea}{\end{eqnarray}}
\newcommand{\ba}{\begin{array}}
\newcommand{\ea}{\end{array}}
\newcommand{\ee}{\end{equation}}
\newcommand{\bes}{\begin{equation*}}
\newcommand{\beas}{\begin{eqnarray*}}
\newcommand{\eeas}{\end{eqnarray*}}
\newcommand{\bas}{\begin{array*}}
\newcommand{\eas}{\end{array*}}
\newcommand{\ees}{\end{equation*}}
\newcommand{\nn}{\nonumber}
\numberwithin{equation}{section}
\begin{document}
\onehalfspacing
\noindent
\begin{titlepage}
\hfill
\vbox{
    \halign{#\hfil         \cr
 \cr
                      } 
      }  
\vspace*{20mm}
\begin{center}
{\Large {\bf    Holographic Complexity for Time-Dependent Backgrounds }\\
}

\vspace*{15mm} \vspace*{1mm} {Davood Momeni$^{a,\ast}$, Mir Faizal$^{b,c,\ddagger}$, Sebastian Bahamonde$^{d,\dagger}$,    Ratbay Myrzakulov$^{a}$}

 \vspace*{1cm}
{\it $^{a}$Eurasian International Center for Theoretical Physics and
Department of General Theoretical Physics, Eurasian National
University, Astana 010008, Kazakhstan\\
 $^{b}$ Irving K. Barber School of Arts and Sciences, University of British
Columbia - Okanagan,  3333 University Way, Kelowna, British Columbia V1V 1V7, Canada \\
$^{c}$Department of Physics and Astronomy, University of Lethbridge, Lethbridge, Alberta, T1K 3M4, Canada.\\
$^{d}$Department of Mathematics,University College London,
Gower Street, London, WC1E 6BT, UK
}

\vspace*{.4cm}

{E-mails : {\tt  $^{\ast}$davoodmomeni78@gmail.com,$^{\ddagger}$mirfaizalmir@googlemail.com,$^{\dagger}$sebastian.beltran.14@ucl.ac.uk}}%

\vspace*{2cm}
\end{center}
\begin{abstract}   
In this paper, we will analyse the holographic complexity for time-dependent asymptotically $AdS$ geometries. 
We will first use a  covariant zero mean curvature slicing of the time-dependent bulk geometries, and then use this 
  co-dimension one spacelike slice of the bulk spacetime to define  
a co-dimension two minimal surface. The time-dependent holographic complexity will    be defined using the volume enclosed by 
  this  minimal surface.  
This time-dependent holographic complexity will reduce to the usual holographic complexity 
for static geometries.  We will analyse the time-dependence as a 
    perturbation of  the  asymptotically $AdS$ geometries. Thus, we will obtain   time-dependent asymptotically $AdS$ geometries, and we will 
   calculate the holographic complexity for such a time-dependent geometries. 
\end{abstract}

\end{titlepage}
\section{Introduction}

An observation made from different branches of physics is that the physical laws can be represented by 
informational theoretical processes  \cite{info, info2}.  The information theory deals with the ability of an observer to process 
relevant information,    it is important to know how much information is lost during a process, and this is 
quantified by   entropy. 
As  the laws  of physics are represented by informational theoretical processes,
entropy is a very important 
physical quantity, and so it   has been used from condensed matter physics to gravitational physics.  
In fact,  in  the   Jacobson formalism,  the geometry of spacetime can also be obtained from the  scaling behavior 
of the maximum entropy of a region of space \cite{z12j, jz12}. This scaling behavior of maximum entropy of a region of space is obtained 
from the physics of black holes. 
As black holes are maximum entropy objects, and this maximum entropy of a black hole scales with its area, 
 it can be argued that 
 the   maximum entropy of a region of space scales with the area  of its boundary.  
 This   scaling behavior has led to the development of the holographic principle   \cite{1, 2}, which states that the degrees of freedom in a region of space 
 is equal to the degrees of freedom on the boundary of that region.   The $AdS/CFT$ correspondence is  a concrete 
  realizations of the holographic principle \cite{M:1997}, and  it relates the string theory  in the bulk of an $AdS$ spacetime to the superconformal field theory on 
  its boundary. 

It is interesting to note that the  holographic principle which  was  initially 
motivated from the physics of black holes, has in turn been used to propose a solution to the black hole information paradox 
\cite{4,5}. 
This is because   it has been proposed that the black hole information paradox might get solved by 
analyzing the of microstates of a black hole, and the 
quantum entanglement of a black hole can be used to study such microstates. This 
quantum entanglement is  quantified using  the holographic
entanglement entropy, and $AdS/CFT$ correspondence can be used to calculate  the holographic
entanglement entropy.  The holographic entanglement entropy 
 of a conformal field theory 
 on the boundary of an asymptotically $AdS$ spacetime  is dual to the area of a minimal surface 
 in the bulk of an asymptotically $AdS$ spacetime.
Thus, for a   subsystem $A$ (with its complement), it is possible to define  $\gamma_{A}$   
as  the $(d-1)$-minimal surface extended 
into the bulk of the $AdS$ spacetime, such that its boundary is $\partial A$. 
This minimal surface is obtained by first foliating the bulk spacetime by constant time slices, 
and then defining a minimal surface on such a slice of the bulk spacetime.  
The holographic entanglement entropy can be calculated using the area of this minimal surface  \cite{6, 6a}
\begin{equation} \label{HEE}
 {S}_{A}=\frac{Area(\gamma _{A})}{4G_{d+1}}\,,
\end{equation}
where $G$  is the gravitational constant for the $AdS$ spacetime. This   minimal surface
is a co-dimension two surface in the bulk spacetime
because of being a co-dimension one submanifold of a particular leaf of the spacelike foliation.
It is possible to generalize the holographic entanglement entropy  to time-dependent geometries  \cite{Hubeny:2007xt}. 
This is because even thought it is not possible to foliate a bulk time-dependent  geometry by a  preferred time slicing, 
 it is possible to foliation a time-dependent  asymptotically $AdS$ geometry  by zero
mean curvature slicing. Thus, it is possible to take slices of the bulk geometry with 
  vanishing trace of extrinsic curvature. This corresponds to taking the 
spacelike slices with maximal area through the bulk, anchored at the boundary. 
This covariant foliation    reduces to the constant time foliation  
 for static geometries.   Thus, a co-dimension one spacelike foliation of time-dependent  asymptotically
 $AdS$ geometry can be performed, 
 and on such a spacelike slice the metric is spacelike, and so a co-dimension two minimal 
 surface can be defined on such a spacelike slice. 
 So, in this formalism, first a maximal spacelike slice 
of the bulk geometry is obtained though the mean curvature slicing,
and then a minimal  surface   $\gamma_{At}$ is constructed on this spacelike 
slice \cite{Hubeny:2007xt}. 
This minimal surface reduces to the usual minimal surface for static geometries, and so for static geometries, 
$\gamma_{At} = \gamma_A$, as the mean curvature slicing reduces to the constant time slicing for such geometries. 
This minimal area surface $\gamma _{At}$ can be  used to define  the time-dependent  holographic entanglement entropy 
for a time-dependent  geometry  \cite{Hubeny:2007xt}, 
\begin{equation} \label{HEE}
{S}_{A}=\frac{Area(\gamma _{At})}{4G_{d+1}}\,. 
\end{equation}
It may be noted that for the static case this time-dependent holographic entanglement entropy 
reduces to the usual definition of holographic entanglement 
entropy, so for the static case, we have ${S}_A (\gamma _{At})= {S}_{A}(\gamma _{A})$. 

The entropy measures how much information is lost in a system. However, it is also important to know  how easy is 
for an observer to obtain  the information present in a system. This  difficulty to obtain information from a system 
is quantified by a new quantity called complexity, just as the loss of information is quantified by entropy. 
  Furthermore, 
as  physical laws can be represented by information theoretical processes, 
 complexity is expected to become another  fundamental physical quantity  describing the  laws of physics. 
 In fact, complexity has already   been used to study condensed matter systems \cite{c1, c2},  molecular physics \cite{comp1}, and   
  quantum computing \cite{comp2}. It is also expected that complexity might  be used to solve the  black hole information paradox, 
  as the recent studies seem to indicate  that the information may not be actually   lost in a black hole, but it would be 
effectively lost,  as it would be impossible to reconstruct it from the Hawking radiation \cite{hawk}. 
However, unlike entropy, there is no universal definition of complexity  of a   system, and there are different proposals for 
defining the complexity of a system.  
It is possible to define the 
complexity of a boundary theory,  as a quantity which is  holographically dual to   a volume of  co-dimension one time 
slice in an anti-de Sitter (AdS) spacetime 
\cite{Susskind:2014rva1,Susskind:2014rva2,Stanford:2014jda,Momeni:2016ekm}. 
In fact, it is possible to used the volume $V(\gamma_A)$ enclosed by the   minimal surface $\gamma_A$ to define holographic 
complexity \cite{Alishahiha:2015rta}. This is the same minimal surface which  was 
used to calculated the holographic entanglement entropy. Thus, we can write the holographic complexity 
as \cite{Alishahiha:2015rta}
\begin{equation}\label{HC}
 \mathcal{C}_A = \frac{V (\gamma_A)}{8\pi R G_{d+1}},
\end{equation}
where $R$ and $V$ are the radius of the curvature and the volume in the $AdS$ spacetime, respectively.   
It may be noted that there are other ways to define the volume in bulk $AdS$, and these 
correspond to other proposals for the
complexity of the boundary theory  \cite{r5}.  It has been possible to use an  alternative proposal  for  
  holographically analyse quantum phase transitions   \cite{r6,r7, r8}. 
However, we shall not use such proposals in this paper, and 
we will only concentrate on the proposal where the  holographic complexity is dual to the volume 
enclosed by the   minimal surface used to calculate the holographic 
entanglement entropy \cite{Alishahiha:2015rta}. 

As we will be analyzing time-dependent  geometries in this paper, we need to generalize 
 holographic complexity to  time-dependent  holographic complexity. It may be noted 
 the holographic entanglement entropy has been generalized to a time-dependent  holographic entanglement entropy using 
 a covariant formalism \cite{Hubeny:2007xt}. Motivated by this definition of time-dependent  holographic entanglement 
 entropy \cite{Hubeny:2007xt}, we will use the same covariant formalism to define the  time-dependent  holographic complexity 
 for time-dependent  geometries.  
 Thus, we will first  foliate  the  time-dependent  asymptotically $AdS$ geometry  by zero
mean curvature slicing, and so each of these  slices of the bulk geometry will have  
  vanishing trace of extrinsic curvature. This will corresponds to taking the 
spacelike slices with maximal area through the bulk, anchored at the boundary. 
Thus, we will get a co-dimension  one surface with a spacelike metric, and  we will again define a  co-dimension two 
minimal surface $\gamma_{At}$ on this spacelike slice of the bulk geometry.  It will be the same minimal surface which was used 
to calculate the time-dependent holographic entanglement entropy \cite{Hubeny:2007xt}. 
 However, now we will calculate the volume 
enclosed by this minimal surface  $V(\gamma_{At})$, 
and use this volume to define the time-dependent  holographic complexity as 
\begin{equation}\label{HC}
 \mathcal{C}_A = \frac{V (\gamma_{At})}{8\pi R G_{d+1}},
\end{equation}
 where $\mathcal{C}_A$ is the time-dependent  holographic complexity. It may be noted that 
 this surface $\gamma_{At}$ reduces to the usual minimal surface $\gamma_A$ for the static geometries, so the volume 
 enclosed by this surface   will also reduce to the volume enclosed by the usual minimal surface for static geometries. 
Thus, this time-dependent  holographic complexity will also reduce to the usual definition of holographic complexity 
 for static geometries, and so for static geometries, we have $\mathcal{C}_A (\gamma_{At})= \mathcal{C}_A(\gamma_{A})$. 
 It may be noted that 
non-equilibrium field theory  has  been used  for analyzing various aspects of the  holography 
using $AdS/CFT$ correspondence, and such study is relevant for holographically 
analysing the time-dependent geometries   \cite{CY, MKT, NNT, CKPT, BDST}. 
 Furthermore,     time-evolution of holographic entanglement entropy 
 has been studied  using the   metric perturbations
 \cite{Kim:2015rvu}. This was done by analyzing the time-dependence as a perturbation of a background geometry. 
 A time-dependent background induced by quantum quench was analysed using the    continuum version of the 
 multi-scale entanglement renormalization    
 \cite{qu}. The 
 causal wedges associated with a given sub-region in the boundary of a time-dependent asymptotically $AdS$ geometry have been used 
 for  understanding  causal holographic information \cite{uq}. This  was done by using a Vaidya-$AdS$ geometry for analysing 
 the  behavior a   null dust collapse in an asymptotically $AdS$ spacetime. 
 In this analysis, the   behavior of holographic entanglement entropy was 
 also discussed.  Holographic complexity, just like holographic entanglement entropy, is a important physical quantity which can 
 be calculated holographically. Therefore, we    have  generalize holographic complexity to 
 time-dependent  holographic complexity,  and now we can use it for  analysing time-dependent geometries.  
So, in this paper, we will analyse the time-dependent holographic complexity for such time-dependent geometries.

\section{Time-Dependent Geometry}\label{timeAdS3}
In this section, we will analyse   a time-dependent asymptotically $AdS$ geometry by analysing   
time-dependent  perturbation of a pure $AdS$ geometry. We will also study the behavior of 
the time-dependent holographic complexity for such a geometry.   
This  time-dependent geometry  can be modeled  using the  Vaidya spacetime, and  
 the metric  for this spacetime   can be written as
\begin{eqnarray}
&&ds^2=\frac{1}{z^2}\Big[-F(t,z)dt^2-2dtdz+H(t,z)dx^2\Big]\,,\label{metric}
\end{eqnarray}
where $F(t,z)$ and $H(t,z)$ are functions of the ingoing Eddington-Finkelstein time coordinate $t$, and $z$ is the radial Poincare
direction. For the specific case where $z = 0$, we recover the $AdS$ boundary.
It is not possible to define a temperature for  a time-dependent  backgrounds as this geometry does not have a time Killing vector. 
However, it  has been demonstrated that the time-dependence  can be analysed as a perturbation around this static geometry, and 
this was done for analysing the time-dependence  of holographic entanglement entropy \cite{ah,Kim:2015rvu}. 
So, we will also analyse the  the time-dependence of this metric as a perturbation around a static geometry, and use it for 
analyzing the time-dependence  of holographic  complexity. 
Now if we 
  we neglect the time-dependence of this geometry, 
  by defining a static geometry with $F(z) = F(z, t)|_{t =0}$, then for this 
  static geometry,  the standard Hawking-Bekenstein 
horizon temperature $T$ can be obtained by 
choosing   the event horizon as the smallest root of the equation $F(z)=0$. 
We will assume
that we have a strip geometry such that its width is  $2L$ in the $x$ direction.  
Now because of   the symmetries of the surface,  $t=t(x)$ and $z=z(x)$ are only functions of $x$, and the   
  surface $\gamma _{A}$ will be characterized by the embedding
\begin{eqnarray}
&&\gamma_{A}=\{t=t(x), z=z(x)\}\,.
\end{eqnarray}
Now for  a   extremal surface which extends smoothly into the bulk, we can assume that the center of the strip
is located at $x=0$. This surface is smooth, and it satisfies the following      the boundary conditions  
\begin{eqnarray}
t(x=0)&=&t^{*}\,, \nonumber \\
z(x=0)&=&z^{*}\,,\\\nonumber
t'(x=0)&=&z'(x=0)=0\, ,  
\end{eqnarray}
 where  prime denotes differentiation with respect to $x$. These boundary conditions define the turning point of 
the strip at $x=0$. It is important to note that the time $t$ in the metric (\ref{metric}) refers to the ingoing time in the
Eddington-Finkelstein coordinates. Since we are located at the  boundary, the physical time   
  is $T = t + z$ (near $z \to 0$). 
Furthermore, at  $x=L$, we need to deal with  the following UV boundary conditions,
\begin{eqnarray}
&&t(x=L)=T-\epsilon\,,\  \ z(x=L)=\epsilon\,. 
\end{eqnarray}
where  $\epsilon\sim 0$ is a cut-off introduced to deal with     
the UV divergence    at the  
boundary $z=0$. 
Now,   we can express the area of the minimal surface  
$\gamma_A$ as 
\begin{eqnarray}
&&\textrm{Area}(\gamma_A)=\int_{-L}^{L}\frac{dx\,H(t(x),z(x))}{z(x)^2}\sqrt{H(t(x),z(x))^2-F(t(x),z(x))t'^2-2t'(x)z'(x)}\,, 
\end{eqnarray}
where   $t=t(x)$ and $z=z(x)$, and so the 
functions $H(t,z)$ and $F(t,z)$ only depend on $x$. It may be noted that 
the Lagrangian density in the integrand has a conserved charge. The area can be now expressed  as
\begin{eqnarray}
&&\textrm{Area}(\gamma_A)=\int_{-L}^{L}dx\Big(\frac{H^2}{z^2}\frac{z^{*}}{H^{*}}\Big)^2\,, 
\end{eqnarray}
where  $H^{*}=H(t^{*},z^{*})$ is a constant. 
Finally, the time-dependent holographic complexity for the metric (\ref{metric}), can be written as
\begin{eqnarray}
&&
\mathcal{C}_A(T)= \lim_{\epsilon\to 0}\frac{V(\gamma_A)}{8\pi R G_{3}}\,,\label{CT}
\end{eqnarray}
where  the co-dimension one volume $V(\gamma_A)$, can be expressed as
\begin{eqnarray}
&&V(\gamma_A)=\int_{T-\epsilon}^{t^{*}}\frac{dt}{t'(t)}\frac{tz'(x(t))}{z(x(t))}\,.
\end{eqnarray}
The quantity (\ref{CT}) is the time-dependent extension of the usual  holographic complexity, which can be used 
for analyzing this time-dependent geometry. 
\par
\begin{figure}[ht]
\centering
     \includegraphics[width=80mm]{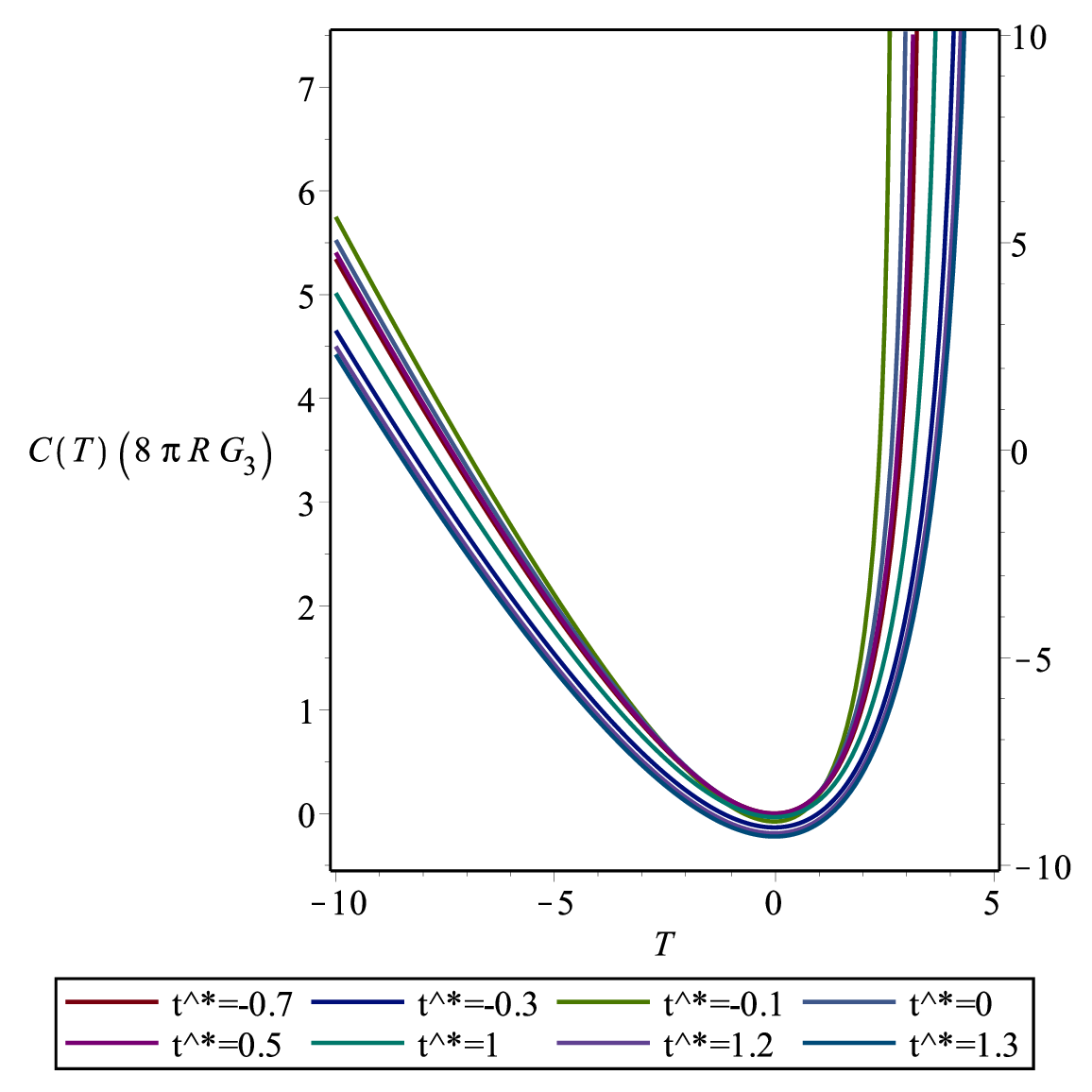}
      \caption{Regularized holographic complexity for pure $AdS_3$ v.s time coordinate $T$.  We used different values for $t^{*}$.
       The schematic graph predicts a minima at $T=0$.  The system at early 
      time has the minimum  for  holographic complexity. }
       \label{constlat}
\end{figure}

The holographic complexity for     a pure $AdS_3$ spacetime can be obtained   by integrating (\ref{CT}), 
 \begin{eqnarray}
&&\mathcal{C}_A(T)=\frac{1}{8\pi R G_3}\Big[t^{*}-T+(t^{*}+z^{*})\log\Big(\frac{z^{*}}{t^{*}+z^{*}-T}\Big)\Big]\,.\label{CT-ADS}
\end{eqnarray}
We will  use numerical analysis  to study the behavior of this  quantity. To do that we will  fix the strip size to be
 \begin{align}
2L=\int_{\epsilon}^{z^{*}}\frac{dz}{z'(z)}\,,
\end{align}
 and then we can find the minimal surface at each  $z^{*}$. We will also     assume  
 $z^{*}\approx 3.39 L$.   Using the 
 Euler-Lagrange equations, we can directly obtain   $z'(z)=\sqrt{\big(z^{*}/z\big)^4-1}$. Fig. \ref{constlat} shows the behavior 
 of the regularized holographic complexity $\mathcal{C}_A(T)$ as a function of the time coordinate $T=t+z$, 
 for different values of $t^{*}$. It can be observed that this has a minima at $T=0$.

Now  from  (\ref{CT-ADS}) and the Fig.\ref{constlat}, we can observe  that the holographic complexity becomes small near $T\approx0$. 
In fact,  near this point, the holographic complexity 
has a minimum. We  observe  that by increasing 
the value of $t^{*}$, the dip   becomes deeper and steeper, for $L =1$.
Thus,    the system will remain close to the $AdS$ boundary $z=0$ if $T=t^*$. In this limit, we observe 
  that the system evolves to a equilibrium state with  $\mathcal{C}_A(T)=0$. Furthermore, at late times, the  tip
point $z^{*}$,  will also meet the horizon.

\section{Metric Perturbations}\label{sec3}
In this section, we will use metric perturbation to analyse to calculate holographic complexity for  time-dependent geometries. 
It may be noted that time-dependent asymptotically $AdS$ geometries are interesting, and have been used for 
analyzing various different systems. In fact, it is expected that using the $AdS/CFT$ correspondence these background will be dual 
to interesting field theory solutions, and these field theory solutions can represent interesting physical systems. In fact, it has been 
demonstrated that 
supergravity solution on a  time-dependent orbifold background is dual to 
a noncommutative field theory with time-dependent noncommutative parameter \cite{accd1}. The 
time-dependent noncommutativity can have lot of interesting applications for 
modeling nonlinear phenomena in quantum optics \cite{accd2}.  
Thus, it is interesting to analyse such backgrounds. Furthermore, we will analyse 
the holographic complexity for such backgrounds, and this can be used to obtain the complexity for the boundary theory. 
Complexity being an important physical quantity in  condensed matter systems \cite{c1, c2} and   molecular physics \cite{comp1}, 
it would be 
interesting to obtain it for condensed matter systems dual to such time-dependent backgrounds.
However, a problem with this approach is that 
the absence of a time Killing vector in such geometries makes it   difficult to perform such an analysis. 
So, in this section, we will use a perturbative technique, which has been used for analysing the 
holographic entanglement entropy of   time-dependent $AdS$ geometries~\cite{Kim:2015rvu}, for analyzing the holographic complexity 
of   time-dependent $AdS$ geometry.  
  The holographic entanglement entropy has been calculated using  a perturbative technique for 
a small deformations of a $AdS$ vacuum spacetime~\cite{Kim:2015rvu}. 
Thus, a background $AdS$ spacetime can  used for this analysis, and on this background spacetime small time-dependent 
perturbation can analysed. The use of  perturbative technique makes it possible to calculate different physical quantities 
on this background. Thus, using the same formalism as has been used for analyzing the time-dependent 
holographic entanglement entropy for a time-dependent 
background, we will now  analyse the time-dependent holographic complexity for such a background. 
Now for  an $AdS_{d+1}$ spacetime with radius $\ell$, the metric  can be written as 
\bea
ds^2=\frac{\ell^2}{\cos^2x}\Big[-dt^2+dx^2+\sin^2 x(d\theta^2+\sin^2\theta d\Omega_{d-2}^2)\Big]\,,
\label{ads}
\eea
where the metric for the unit-sphere is denoted by $d\Omega_{d-2}^2$. Here we have 
  split the $S^{d-1}$ metric into a polar angle $\theta$ 
and a unit-radius sphere $S^{d-2}$. 

Now, we will assume that an  entangling region on the boundary is a cap-like one defined between
$0\le \theta <\theta_0$, and  we  will    use  
the boundary time $t=t_0$. 
The extremized area, which can be used  to obtain 
the holographic complexity, is given by the following functional of $x(\theta)$,
\bea
\textrm{Area}=\ell^{d-1}\text{vol}(S^{d-2})\int^{\theta_0}_0
d\theta\frac{( \sin x\sin\theta)^{d-2}}{\cos^{d-1}x}\sqrt{\left(\frac{dx}{d\theta}\right)^2+\sin^2 x} \,\, .   \label{area}
\eea
Now if we assume  that $\theta_0<\pi/2$, then the  holographic complexity can be written as 
\bea
\mathcal{C}_A=\frac{\ell^{d-1}\text{vol}(S^{d-2})}{8\pi G }\int^{\theta_0}_0 d\theta(\sin\theta)^{d-2} \int_{0}^{x(\theta)} dx\frac{( \sin x)^{d-1}}{\cos^{d}x}\,\, .
\label{V}
\eea
It is important to mention that it is   difficult  to find the most general exact solution for $x(\theta)$ derived from the action 
\eqref{area}. However, we can use an appropriate solution which satisfies the Euler-Lagrange equations in  any dimension $d$, and we 
will use this  solution to analyse metric perturbations. Further, we will assume that at the boundary,
this solution satisfies    $x=\pi/2$, 
and it is mapped to $a=\cos\theta_{0}$. We will call this solution as the   constant-latitude solution, and it will be explicitly
written as 
\bea
x(\theta)=\sin^{-1} \left(\frac{a}{\cos\theta}\right) \,. 
\label{cl}
\eea
For this solution, from  (\ref{V}), we obtain  
\bea
\mathcal{C}_A=\frac{\ell^{d-1}\text{vol}(S^{d-2})}{8\pi G (d-1)}\int^{\theta_0}_0 d\theta(\sin\theta)^{d-2}\big(1-\frac{a^2}{\cos^2\theta}\big)^{1-d}F\Big(\frac{1-n}{2},
1-\frac{n}{2},\frac{3-n}{2},1-\frac{a^2}{\cos^2\theta}\Big) \,.
\label{C2}
\eea
Now we can    investigate 
small deformations of an $AdS$ background and then analyse the  corrections to the 
minimal area solution \eqref{cl} from those small deformations.  These corrections will produce correction terms for the volume 
enclosed by this minimal surface, and this will in turn produce  correction terms for  the holographic complexity. 
We can parametrize the  coordinates $\theta$ and $x$ as 
\bea
z =\cos\theta, \quad\rho =\sin x \, . 
\eea
Now   the area functional can be written as 
\bea
\text{Area}=\ell^{d-1}\text{vol}(S^{d-2})\int^1_{\cos^{-1}\theta_0} dz 
\frac{\rho^{d-2}(1-z^2)^{(d-2)/2}}{(1-\rho^2)^{d/2}}\sqrt{(\rho')^2+\frac{\rho^2(1-\rho^2)}{1-z^2}}\,.
\label{area2}
\eea
Furthermore, using  this parametrization, the   holographic complexity can be  expressed as
\bea
\mathcal{C}_A=\frac{\ell^{d-1}\text{vol}(S^{d-2})}{8\pi G (d-1)}\int_{\cos^{-1}\theta_0}^1 dz  (1-z^2)^{\frac{d-3}{2}}
\big(1-\frac{a^2}{z^2}\big)^{1-d}F\Big(\frac{1-n}{2},1-\frac{n}{2},\frac{3-n}{2},1-\frac{a^2}{z^2}\Big)\,.
\label{C21}
\eea
In order to calculate the holographic complexity, we need to substitute the solution $\rho=a/z$ back into \eqref{C21},
and then perform the integral. 
As is common in $AdS/CFT$ correspondence, this 
  integral is divergent near the $AdS$ boundary,  $x=\pi/2$. So, 
 we   introduce the following   cut-off
 \begin{align}
 x_m=\pi/2-\epsilon\, . 
 \end{align} 
Now by  mapping the original solution \eqref{cl}, we obtain 
\bea
\theta_{m}=\theta_0-\frac{1}{2}\epsilon^2\cot\theta_0 ,\quad \text{i.e.} \quad z_{m}=
a\left(
1+\frac{\epsilon^2}{2}
\right)
\,.
\eea
 
It is well-known that excited states in conformal field theory tare dual to deformations of the $AdS$ spacetime. So, it is interesting to analyze 
small metric perturbations around an $AdS$ spacetime. These small deformation can be used to obtain the 
  holographic complexity for excited states of the dual conformal field theory.  As we want to analyse the time-geometries, we will analyse 
  the time-dependent deformations. As this     spacetimes  will also be  spherically symmetric, we can  write the metric for this spacetime as      
\bea
ds^2=\frac{\ell^2}{\cos^2x}\biggl(-A(t, x)e^{-2\delta(t, x)}dt^2+A^{-1}(t, x)dx^2+\sin^2x(d\theta^2+\sin^2\theta d\Omega^2_{d-2})\label{metric3}
\biggr) \, ,
\eea
where the pure $AdS$ metric is recovered by choosing $A=1$ and $\delta=0$. 
In terms of new variables $\rho=\sin x, z=\cos \theta$,  and    $\rho=\rho(z)$ and $t=t(z)$, 
the area functional \eqref{area2}  can   be expressed as 
\bea
\frac{\text{Area}}{\text{vol}(S^{d-2})}=\int dz\frac{\rho^{d-2}(1-z^2)^{\frac{d-3}{2}}}{(1-\rho^2)^{\frac{d-1}{2}}}
\sqrt{-g_{tt}(1-z^2)(t')^2+\frac{g_{xx}(1-z^2)}{1-\rho^2}(\rho')^2+\rho^2}\,,
\label{the area2}
\eea
where primes denote differentiation with respect to $z$, and  
\bea
g_{tt}&=&Ae^{-2\delta}=1-2\sum_{n=1}^\infty\nu_{n}(t, \rho)\epsilon^{n}\,,
\\
g_{xx}&=&A^{-1}=1+2\sum_{n=1}^\infty m_{n}(t, \rho)\epsilon^{n}\,.
\eea
Here, we have also defined a dimensionless  perturbation parameter $\epsilon$. 
Now from  Eq. (\ref{metric3}), we obtain that the holographic 
complexity  for such a time-dependent geometry as, 
\bea
&&\mathcal{C}_A=\frac{\text{vol}(S^{d-2}) l^{d-1}}{8\pi G}\int d\rho\frac{\rho^{d-1}}{\sqrt{1-\rho^2}}\int dz\frac{A^{-1/2}(t(z),\rho)z^{d-2}}{\sqrt{1-z^2}}\,.
\label{HC4}
\eea
We can also use a perturbative approach to compute the 
minimal-area surface. We know that for  the pure $AdS$ spacetime, 
the solutions are $\rho(z)=a/z,t(z)=t_0$, and so,  we can assume that the 
perturbative solutions will statisfy 
\bea
\rho(z)&=&\frac{a}{z}+\sum_{n=1}^\infty\rho_{n}(z)\epsilon^{n}\,, 
\\
t(z)&=&t_0+\sum_{n=1}^\infty t_{n}(z) \epsilon^{n}\,,
\eea
where we have expanded the solutions in term of the small parameter $\epsilon$. So, we  can analyse such 
solutions using this perturbative technique. We can thus obtain time-dependent holographic complexity for different 
time-dependent geometries. 

\section{Deformation}\label{sec4}
In this section, we will analyse the time-dependent holographic complexity for  deformations of  $AdS$ spacetime. 
It may be noted that various deformation of the $AdS$ spacetime are  dual to interesting physical systems. 
It has been demonstrated that a 
Schwarzschild black hole in an $AdS$ background  can be used to analyse   high spin baryon in hot strongly coupled plasma. 
This is because such  a system can be analysed  using 
the finite-temperature supersymmetric Yang-Mills theory, 
and this theory is dual to the Schwarzschild black hole in a  $AdS$ background  \cite{accd4}. 
So, it would be interesting to analyse such deformations of a time-dependent $AdS$ background, as this can be used 
to obtain the complexity of the field theory dual to such backgrounds. 
It may be noted that   such a deformation of a time-dependent $AdS$ spacetime has been 
used to obtain the conformal field theory dual to a  FLRW background \cite{accd5}. It would be interesting to analyse the holographic 
complexity for such systems, as this is an important physical quantity. So, in this section, 
we analyse a  simple examples of a deformation $AdS$ spacetimes, and then we evaluate the integral (\ref{HC4}) using  perturbative
 techniques. Thus, we will find the first order corrections 
to the minimal surfaces $ \rho_{1}(z),t_{1}(z)$, and then  use it to obtain the corrections to   
holographic complexity.  
Now we can analyse  a geometry with a small mass, which is a minimal deformation of the pure $AdS$ spacetime.
The mass terms deforms this  geometry   to a light 
Schwarzschild-$AdS$  spacetime.  
  Thus, we will use this time-dependent formalism for  analysing the time-dependent 
Schwarzschild black hole in $AdS$ background. 
The horizon, which is located at $r=h$,  satisfies $f_d(h)=0$,
and so we can    define the following perturbative parameter, 
\begin{align}
\epsilon\equiv M \ell^2/h^d\ll1\,.
\end{align}
The area  expressed in terms of $t(\theta)$ and $ r(\theta)$,  can be written as 
\bea
\text{Area}=\text{vol}(S^{d-2})\int_{0}^{r_0}dr (r \sin\theta(r))^{d-2}\sqrt{\frac{1}{f}-ft'^2+r^2\theta(r)'^2}\,,\label{areasads}
\eea 
where prime denotes differentiation with respect to $r$ , and the extremal surfaces are defined by $t(r)$ and $\theta(r)$. 
The holographic complexity in usual coordinates can be written as 
\bea\label{C3}
\mathcal{C}_A=\frac{\text{vol}(S^{d-2})t_0}{d}\int_{0}^{\theta_0}r(\theta)^d(\sin\theta)^{d-2}d\theta\,.
\eea 
In order to evaluate the integral, we have to obtain the solution to the 
  Euler-Lagrange equation for $\theta(r)$. As we want to apply this formalism to a specific example, so we will now 
 apply it to $AdS_3$ to simplify calculations. The $AdS_3$ spacetime has been used to analyse various interesting physical 
 systems. The   holographic duals to time-like warped $AdS_3$ spacetimes have been studied, and it was demonstrated that such systems 
 have at least one    Virasoro algebra with computable central charge \cite{viro}. In fact, it was also observed that 
 there exists a dense set of points in the moduli space of these models in which there is also a second commuting Virasoro algebra. 
 The higher spin theories on an $AdS_3$ background have   been studied \cite{dsig}. The field theory  dual to such a background  has also been analysed, 
 and constraints on the     central charge of such a field theory dual have been obtained from  the modular invariance. 
   It may be noted that time-dependent solution for D-branes have been analysed using $AdS_3$ spacetime \cite{dbran}. 
   In this work, D-branes solutions where analysed using a $\kappa$-deformed background with non-trivial dilaton and Ramond-Ramond fields. 
   So, it is interesting to analyse the time-dependent deformation of $AdS_3$ spacetime. 
 The $AdS_3$ spacetime has also been used to analyse the microstates of black holes \cite{blacw}.  
 Thus, $AdS_3$ spacetime has been  used for  analyzing  interesting physical systems, and it would be interesting to analyse the 
 deformation of $AdS_3$ spacetime. 
 So,   we will apply this formalism to 
 $AdS_3$ spacetime, and the  equation for $\theta(r)$     for this spacetime can be written as 
(\ref{areasads}). 
Now for this spacetime geometry, we obtain 
\bea
\frac{r^2\theta'(r)}{\sqrt{\frac{1}{f}-ft'^2+r^2\theta(r)'^2}}=p,\ \ \frac{-ft'}{\sqrt{\frac{1}{f}-ft'^2+r^2\theta(r)'^2}}=E.
\eea 
As  this metric is static, we can write 
$g_{ta}=0$ and $\partial g_{\mu\nu}/\partial t=0$. Thus, the equation for $t(z)$  at leading  order
gives us a trivial solution $t^{0}(z)=t_0$,  and the equations for $t(r)$ and $\theta(r)$ can be written as 
\bea
{\frac {d}{dr}}t \left( r \right)&=&\displaystyle{\frac {{\ell}^{3}Er \left( 
\epsilon\,{h}^{2}{p}^{2}-{h}^{2}\epsilon\,{r}^{2}+{\ell}^{2}{r}^{4}-{\ell}^{
2}{p}^{2}+{r}^{2}{\ell}^{2}-{r}^{2}{\ell}^{2}{p}^{2}+{E}^{2}{\ell}^{2}{r}^{2}
 \right) ^{-1/2}}{-{\ell}^{2}-{r}^{2}{\ell}^{2}+\epsilon\,{h}^{2}}}\,,\\
{\frac {d}{dr}}\theta \left( r \right) &=&\displaystyle{\frac {p \left( 
\epsilon\,{h}^{2}{p}^{2}-{h}^{2}\epsilon\,{r}^{2}+{\ell}^{2}{r}^{4}-{\ell}^{
2}{p}^{2}+{r}^{2}{\ell}^{2}-{r}^{2}{\ell}^{2}{p}^{2}+{E}^{2}{\ell}^{2}{r}^{2}
 \right) ^{-1/2}}{r}}\,.
\eea 
We can  expand $f$ in series of $\epsilon$ as $f=(1+r^2/\ell^2)-\epsilon\, r_{+}^2/\ell^2$.  Using this expansion, we can 
solve the above 
equation for $t(r)=t_0+ \,t^1(r)$ and $ \theta(r)=\theta^{0}(r)+\epsilon \,\theta^{1}(r)$. Thus, up to first order in $\epsilon$, 
we obtain
\bea
&&t^{1}(r)=\frac{1}{2\ell^2}\, \left( -E{h}^{2}\int \!{\frac {3\,{r}
^{5}+ \left( 3-3\,{p}^{2}+2\,{E}^{2} \right) {r}^{3}-3\,r{p}^{2}}{
 \left( {r}^{4}+ \left( 1-{p}^{2}+{E}^{2} \right) {r}^{2}-{p}^{2}
 \right) ^{3/2} \left( 1+{r}^{2} \right) ^{2}}}{dr}+2\,{C_1}\,{\ell}
^{2} \right)\,,\\
&&\theta^1(r)=-\frac{1}{2\ell^3}\, \left( p{h}^{2}\int \!{\frac {{p}^{2}-{r}^{
2}}{r \left( {r}^{2}-{r}^{2}{p}^{2}+{r}^{4}-{p}^{2}+{E}^{2}{r}^{2}
 \right) ^{3/2}}}{dr}-2\,{C_2}\,{\ell}^{3} \right)\,.
\eea
Finally, using  this solution, we    obtain   the holographic complexity for this geometry, 
\bea
&&\Delta\mathcal{C}=\mathcal{C}_{SAdS_3}
-\mathcal{C}_{AdS_3}\approx\frac{t_0}{2}\frac{M \ell^2}{h^2}\int_{0}^{\theta_0}r^2\partial_r\theta^{1}(r)dr. 
\eea
The first order correction for the  holographic complexity of  the $AdS_3$ background can be written as 
\bea
 \Delta\mathcal{C}&=&\frac{t_0}{4}\frac{M \ell^2}{h^2}\, Im\Big[ \Big( \left( 2\,{p}^{2}+1+2\,{E}^{2}
+{p}^{4}-2\,{p}^{2}{E}^{2}+{E}^{4} \right) {\ell}^{3}\Xi\Big)^{-1}\times\nonumber\\
&&\Big(p{h}^{2} \left(i{p}^{
	2}{E}^{2} -i{p}^{2}-i{p}^{4}+i{p}^{2}{r_{{0}}}^{2}+i{r_{{0}}}^{2}+i{E}^{2}{r_{{0}}}^{2}+p
\Xi+{p}^{3}\Xi-p\Xi\,{E}^{2} \right) \Big)
 \Big]\,,
\eea
where we have introduced the parameter $\Xi=\sqrt {{r_{{0}}}^{4}+{r_{{0}}}^{2}-{p}^{2}{r_{{0}}}^{2}+{E}^{2}{r_
{{0}}}^{2}-{p}^{2}}
$ and $Im(z)$ denotes the imaginary part of $z$.

 After we have  demonstrated how this formalism can be used to analyse perturbations, 
 we will apply it for  analysing  a time-dependent  metric. So,
  we will  find the holographic complexity  for a time-dependent deformation of 
the $AdS_3$, given by the metric (\ref{metric3}). 
Thus, using the time-dependent formalism \cite{Kim:2015rvu}, we now analyze  the spectrum of a 
massive scalar field     in the background (\ref{metric3}).
The scalar field   with conformal mass $m^2=\Delta(\Delta-d)/\ell^2$  can be described by the following equation, 
\be
\partial_t (e^{\delta} A^{-1} \partial_t \phi ) - 
\frac{1}{\tan^{d-1}x} \partial_x ( A e^{-\delta} \tan^{d-1} x \partial_x \phi )
+\frac{\Delta(\Delta-d)}{\cos^2 x} e^{-\delta} \phi = 0 \, . 
\label{kge}
\ee
The Einstein's field  equations   reduced to the    first order system of differential 
equations for this  scalar field $\phi$,
\bea
\delta' &=& -  \sin x \cos x ( A^{-2} e^{2\delta} \dot\phi^2 + 
\phi'^2 ) \, , 
\label{ee1}\\
A' &=& A\delta' + \frac{d-2+2\sin^2x}{\sin x\cos x} (1-A) - 
\frac{\Delta(\Delta-d)\sin x}{\cos x} \phi^2 \, . 
\label{ee2}
\eea
We can now  use a perturbative technique to find solutions for this system.  Thus, we will use a  small deformations 
of the metric in the $AdS_3$ spacetime. So, we will use  the following time-dependent background,
\bea
\delta(t, u)&=&\epsilon \left[-1+u^8+\frac{3\cos(8t)}{5}+u^8\cos(8t)-\frac{8u^{10}\cos(8t)}{5} \right]\,,
\\
A(t, u)&=&1-\epsilon \left[\frac{2u^4}{3}+\frac{2u^6}{3}-\frac{4u^8}{3}-2u^8\cos(8t)+2u^{10}\cos(8t)\right] \,.
\eea
where $u=\cos x=\sqrt{1-\rho^2}$. We have to use  a first order perturbation regime for these metric functions,  and the perturbed
functions, $t(z)$ and $\rho(z)$. These solutions  are  given by \cite{Kim:2015rvu},
\bea
\rho_1(z)&=&\frac{a(a^2-z^2)^2(36z^4-a^2z^2(33+25z^2)+2a^4(5+3z^2+3z^4))}{105z^9}\label{rho1}
\nn\\
&&+\frac{a(a^2-z^2)^2}{315z^{11}}\biggl(-82z^6+a^2z^4(151+95z^2)-2a^4z^2(60+31z^2+32z^4)
\nn\\
&&+a^6(35+15z^2+16z^4+16z^6)
\biggr)\cos(8t_0)\,,\\
t_1(z)&=&-\frac{(a^2-z^2)^2}{315z^{10}}\biggl(-19z^6+19a^2z^4(-2+5z^2)+a^4(69z^2-62z^4-64z^6)
\nn\\
&&+a^6(-28+15z^2+16z^4+16z^6)\biggr)\sin(8t_0)\,.
\eea
So, for   $d =2$,   we can evaluate (\ref{HC4}), up to first order in $\epsilon$, 
\bea
&&\mathcal{C}_A=\frac{  \ell}{8\pi G}\int d\rho\frac{\rho}{\sqrt{1-\rho^2}}\int dz\frac{A^{-1/2}(t(z),\rho)}{\sqrt{1-z^2}}\,.
\label{HC41}
\eea
Finally, by perturbing $\delta(t, u)$ and $A(t, u)$ for $t(z)=t_0+\epsilon\, t_1(z)$ and $\rho(z)=a/z+\epsilon\,\rho_1(z)$, we obtain,
the following expression for holographic complexity 
\bea
&&\Delta\mathcal{C}=\frac{\ell}{24\pi G}\int_{\eta}^{1}{\frac {\Pi(z)}{{z}^{11}\sqrt {{z}^{2}-{a}^{2}}\sqrt {1-{z}^{2}}}}\,,\label{deltaC}
\eea
where 
\bea
\Pi(z)&=&-3\,{z}^{12}+3\,a{z}^{11}\rho_{{1}} \left( z \right) +11\,{a}^{4}{z}^{
8}+3\,{z}^{10}{a}^{2}\cos \left( 8\,t_{{0}} \right) -15\,{a}^{4}{z}^{8
}\cos \left( 8\,t_{{0}} \right)+30\,{a}^{6}{z}^{6}\cos \left( 8\,t_{{0
}} \right)\nonumber\\&&  -15\,{a}^{6}{z}^{6}-30\,{a}^{8}{z}^{4}\cos \left( 8\,t_{{0}
} \right) +9\,{a}^{8}{z}^{4}+15\,{a}^{10}\cos \left( 8\,t_{{0}}
 \right) {z}^{2}-2\,{a}^{10}{z}^{2}\nonumber\\
 &&-3\,{a}^{12}\cos \left( 8\,t_{{0}}
 \right)\,.
 \eea
It may be noted that at the boundary $z=a$, 
we have to  introduce a cut-off $\eta$. 
Thus,  Eq. (\ref{deltaC}) can be written as 
\bea
\Delta\mathcal{C}&=&\frac{ \ell}{24\pi G}
\sum _{m=0}^{\infty } \left(  \left( 2\,m \right) !\,{4}^{-m}\sum _{k=0
}^{\infty }{\frac { \left( {a}^{-2} \right) ^{k} \left( 2\,k \right) !
\,{4}^{-k}{z}^{2\,k}}{\sqrt {-{a}^{2}} \left( k! \right) ^{2}}}  \left( m! \right) ^{-2} \right)\times\nonumber\\
&&\Big[B_{mk}
 \left( 1 \right)-B_{mk}
 \left( \eta\to0 \right)\Big]\,,
\eea
where $\rho_1(z)$ is given by (\ref{rho1}),  and we have defined
\begin{eqnarray}
B_{mk}(z)&=&\int dz\, {z}^{2\,m-11+2\,k}( -3\,{z}^{12}+3\,a{z}^{11}\rho_{{1}} \left( z \right) +11\,{a}^
{4}{z}^{8}+3\,{z}^{10}{a}^{2}\cos \left( 8\,t_{{0}} \right) -15\,{a}^{
4}{z}^{8}\cos \left( 8\,t_{{0}} \right)) \nonumber\\ 
&&+\int dz \,{z}^{2\,m-11+2\,k}( 30\,{a}^{6}{z}^{6}\cos
 \left( 8\,t_{{0}} \right) -15\,{a}^{6}{z}^{6}-30\,{a}^{8}{z}^{4}\cos
 \left( 8\,t_{{0}} \right) +9\,{a}^{8}{z}^{4}\nonumber \\
&&+15\,{a}^{10}\cos \left( 
8\,t_{{0}} \right) {z}^{2})- \int dz\, {z}^{2\,m-11+2\,k}( 2\,{a}^{10}{z}^{2}+3\,{a}^{12}\cos \left( 8
\,t_{{0}} \right)  )\,.
\end{eqnarray}
Thus, we are able to calculate the holographic complexity for a time-dependent background. 
This formalism can be used to obtain the complexity of the field theory dual to such a time-dependent background. 
Furthermore, it is also possible to use this formalism to analyse other time-dependent asymptotically $AdS$ geometries. 
So, we can use the results of this paper to analyse holographic complexity for various deformations of the $AdS$ geometries. 

\section{Conclusions}\label{secfinal}
In this paper, we analyse the holographic complexity for time-dependent geometries. 
Just as  the holographic entanglement entropy is dual to an area in the bulk of an $AdS$ spacetime, 
holographic complexity is 
as a  quantity dual to a volume in 
the bulk $AdS$. In this paper, the concept of holographic complexity was generalized to a time-dependent holographic complexity. 
This time-dependent  holographic complexity was defined as a quantity  dual to volume 
of a  region in a time-dependent $AdS$ geometry.   Thus, we   first  foliated  the  time-dependent  asymptotically
$AdS$ geometry  by zero
mean curvature slicing, and  obtained     slices of the bulk geometry with     
  vanishing trace of extrinsic curvature. This   corresponded  to taking the 
spacelike slices with maximal area through the bulk, anchored at the boundary. 
Thus, we obtained a co-dimension one surface with a spacelike metric. 
We used this  co-dimension  one surface to defined a  co-dimension two 
minimal surface in  the bulk geometry.  It was  be the same minimal surface which was used 
for calculating  the time-dependent holographic entanglement entropy \cite{Hubeny:2007xt}. 
 However, we    calculating  the volume 
enclosed by this minimal surface, 
and use this volume to define the time-dependent  holographic complexity. 
It was observed that this definition of time-dependent holographic complexity reduced to the usual 
holographic complexity for static geometries. 
The metric perturbation has been used in this paper to analyse this
behavior of time-dependent complexity. 
This has been motivated by the earlier works on the 
holographic entanglement entropy where such  a perturbative technique was used for analysing the effects of 
 small deformations of a $AdS$ spacetime. Thus, in this paper, the same technique was used for analysing holographic complexity for 
 a small deformation of a $AdS$ spacetime. This formalism was finally applied for analyzing a time-dependent geometry.
 Thus,   holographic complexity for a time-dependent  background was studied in this paper.

It will be interesting to analyse other time-dependent backgrounds using the formalism developed in this paper. 
It may be noted that  the  string theory propagating in a pp-wave time-dependent  background with a null singularity has been studied 
\cite{a1}. 
In this analysis,  it has been demonstrated that 
 entanglement entropy is dynamically generated by this  background.   It would be also interesting to analyse the 
 holographic complexity  for the  string theory propagating in a pp-wave time-dependent  background with a null singularity. 
 Furthermore, the holographic entanglement entropy has been studied for various interesting systems, and it would be interesting 
 to analyse the holographic complexity for such systems. 
The effects of deforming   the renormalized entanglement entropy  
near the UV fixed point of a three dimensional field theory by a mass terms have been studied 
\cite{a}.
This analysis was performed using  the  Lin-Lunin-Maldacena  
geometries corresponding to the vacua of the mass-deformed ABJM theory. Thus, 
 the small mass effect for various droplet configurations were 
analytically compute, and it was demonstrated that   the  
  renormalized entanglement entropy is monotonically decreasing at the UV fixed point. 
  It would be interesting to calculate the holographic complexity 
  for this system, and use it to analyse the behavior of this system.

\section*{Acknowledgments}
We would like to 
thank the referee for useful comments. 
S.B. is supported by the Comisi\'on Nacional de Investigaci\'on
Cient\'ifica y Tecnol\'ogica (Becas Chile Grant No. 72150066).

\end{document}